\documentclass[12pt]{article}

\usepackage{amssymb,amsmath} 

\newcommand{\be}{\begin{equation}}
\newcommand{\ee}{\end{equation}}
\newcommand{\Dlt}{\Delta}

\newcommand{\bt}{\beta}

\newcommand{\ep}{\varepsilon}
\newcommand{\al}{\alpha}
\newcommand{\ra}{\rightarrow}

\newcommand{\gm}{\gamma}

\begin{document}

\begin{center}

{\Large{\bf Self-similar extrapolation from weak to strong coupling} \\ [5mm]
S. Gluzman$^1$ and V.I. Yukalov$^2$} \\ [3mm]

{\it
$^1$Generation 5 Mathematical Technologies Inc., Corporate Headquarters, \\
515 Consumer Road, Toronto, Ontario M2J 4Z2, Canada \\ 
and \\
$^2$Bogolubov Laboratory of Theoretical Physics, \\
Joint Institute for Nuclear Research, Dubna 141980, Russia}

\end{center}

\vskip 3cm

\begin{abstract}

 The problem is addressed of defining the values of 
functions, whose variables tend to infinity, from the knowledge of 
these functions at asymptotically small variables close to zero. 
For this purpose, the extrapolation by means of different types of 
self-similar approximants is employed. Two new variants of such an 
extrapolation are suggested. The methods are illustrated by several
examples of systems typical of chemical physics, statistical physics,
and quantum physics. The developed methods make it possible to find 
good approximations for the strong-coupling limits from the knowledge 
of the weak-coupling expansions.  
\end{abstract}

\vskip 2cm

{\bf Keywords} Weak-coupling expansions. Asymptotic series. 
Extrapolation of asymptotic series. Strong-coupling limit.
Self-similar approximation theory.

\newpage

\section{Introduction}

In a variety of problems of chemistry and physics [1], there is the 
necessity of defining the properties of systems in the strong-coupling 
limit, when, however, this limit cannot be accessed directly, but when, 
because of the complexity of the problem, only the weak-coupling expansion 
is available. Then the principal question is whether one can infer the 
strong-coupling properties from a weak-coupling expansion?  

In the present paper, we address this principal question and show that the
problem can be resolved by employing the {\it self-similar approximation 
theory} [2-12] allowing for the extrapolation of asymptotic series to finite 
values of expansion parameters. Here we concentrate on the question of how
to better organize an extrapolation not merely to finite parameter values,
but to their {\it infinite} values. We show that the self-similar 
extrapolation is a convenient tool for this purpose. Since this approach 
allows for the construction of different approximation schemes, it is 
useful to compare the latter in order to decide which of them better suits 
the posed problem. It is also necessary to make some modifications for 
adjusting the methods to the case, when the behavior at infinity is of 
interest. Here we make such modifications and suggest two new ways of 
constructing self-similar approximants, which have not been analyzed in 
our previous publications.

To illustrate the applicability of the developed methods and to compare 
their accuracy, we consider several problems whose mathematical structure
is typical of chemical physics, quantum physics, and statistical physics.
We compare different variants of the self-similar approximants. Also, we 
consider the Pad\'e approximants [13] and show that the accuracy of 
the best Pad\'e approximant is always inferior to that of the best 
self-similar approximant. In some cases, the Pad\'e approximants 
are not applicable at all, yielding qualitatively wrong results, while 
there always exists an accurate self-similar approximant.

In Sec. 2, we give the general formulation of the problem, defining the
strong-coupling exponents and amplitudes. We introduce two novel methods 
of constructing the self-similar approximants yielding the {\it corrected 
self-similar approximants} and the {\it iterated root approximants}. For 
the beginning, we choose, in Sec. 3, the Debye-H\"uckel function 
from the theory of strong electrolytes. The quartic anharmonic oscillator 
of Sec. 4 provides the weak-coupling asymptotic series with the mathematical 
structure that is common for many problems of quantum mechanics and field 
theory. In Sec. 5, we consider the expansion factor of a polymer chain. 
The ground-state energy for a model of one-dimensional string is calculated 
in Sec. 6. A more realistic model of fluctuating fluid membrane is treated 
in Sec. 7. A dilute gas of strongly interacting fermions is studied in 
Sec. 8. Then, in Sec. 9, the strong-coupling limit of a one-dimensional 
Bose system is analyzed. The problem of calculating the Bose-Einstein 
condensation temperature of an interacting Bose gas is investigated in 
Sec. 10. Finally, Sec. 11 concludes.  
 
Some of the examples, analized in the present paper, have been considered
earlier [14,15], however, by other methods. The main difference of this 
paper from our previous publications is threefold: (i) We consider a large 
variety of problems to show the generality of our theory. (ii) We give a
comparative analysis of different approximants. (iii) We suggest new 
variants of self-similar approximants, which have not been considered 
previously.

\section{Formulation of general problem}

Let us be interested in a real function $f(g)$ of a real coupling parameter 
$g$, when $g \rightarrow \infty$, which corresponds to the strong-coupling 
limit. But assume that the problem is so complicated that we are not able to 
calculate this limit directly. Instead, we can get only the weak-coupling 
expansion, when perturbation theory is applicable. Then, we can find the 
value
\be
\label{1}
 f(g) \simeq f_k(g) \qquad (g \ra 0)  
\ee
for asymptotically small $g$, which is given by the weak-coupling expansion
\be
\label{2}
 f_k(g) = f_0(g) \sum_{n=0}^k a_n g^n \; . 
\ee
Here the factor $f_0(g)$ is not expandable in powers of $g$. Without the 
loss of generality, we can always set $a_0 = 1$, which, however, is not 
principal.

\subsection{Strong-coupling exponents and amplitudes}

>From some additional information, we may know that the behavior of the 
sought function at large coupling parameter is as follows:
\be
\label{3}
f(g) \simeq B g^\bt \qquad (g \ra \infty)  \; .
\ee
The power index $\beta$ is called the {\it strong-coupling exponent} and can 
be defined as
\be
\label{4}
\bt \equiv \lim_{g\ra\infty} \; \frac{\ln f(g)}{\ln g} \;   .
\ee
This can also be written as 
$$
 \bt = \lim_{g\ra\infty} \;
\frac{d\ln f(g)}{d\ln g} = \lim_{g\ra\infty} \;
\frac{g}{f(g)}\; \frac{d f(g)}{dg} \;  .
$$
We assume that this exponent is known, and what we need to find is the
{\it strong-coupling amplitude}
\be
\label{5}
B \equiv \lim_{g\ra\infty} \; \frac{f(g)}{g^\bt}    .
\ee
The problem we confront is how to find the strong-coupling amplitude (5) 
from the knowledge of the weak-coupling expansion (2)?

\subsection{Corrected self-similar approximants}

An effective extrapolation can be done by means of the self-similar 
approximation theory [2-12]. Then, from the weak-coupling expansion 
$f_k(g)$, we construct a self-similar approximant $f^*_k(g)$. When
the behavior of the sought function is supposed to be exponential, 
we have to use the self-similar exponential approximants [16,17]. 
But in the considered case (3) of the power law behavior, it is 
better to employ the self-similar factor approximants [18-21]. 

Suppose, we have found an approximant $f^*_k(g)$. This can be an 
exponential approximant [16,17] or a factor approximant [18-21].  
Now we suggest a novel way of correcting this approximant and, 
respectively, improving the accuracy of calculating the strong-coupling 
amplitude. 

Let an approximant $f^*_k(g)$ be given, whose parameters are defined 
by the accuracy-through-order procedure, which implies that the 
approximant $f^*_k(g)$ is re-expanded in powers of $g$ and compared, 
term by term, with the initial expansion, so that
\be
\label{6}
 f_k^*(g) \simeq f_k(g) \qquad (g \ra 0) \;  .
\ee       
This is also called the {\it re-expansion procedure}. More details, 
describing this procedure, can be found in Ref. [15]. 

In constructing the approximant $f^*_k(g)$, we keep in mind the limiting 
condition
\be
\label{7}
 \lim_{g\ra\infty} \; \frac{\ln f_k^*(g)}{\ln g} = \bt  .
\ee
Our aim is to find the strong-coupling amplitude
\be
\label{8}
 B_k \equiv \lim_{g\ra\infty} \; \frac{f_k^*(g)}{g^\bt}   .
\ee

Now, let us introduce a {\it correcting function}
\be
\label{9}
 C_{k+p}(g) \simeq \frac{f_{k+p}(g)}{f_k^*(g)} \qquad
(g \ra 0) \; ,
\ee
which is defined as an expansion in powers of $g$, such that
\be
\label{10}
 C_{k+p}(g) = \sum_{n=0}^{k+p} b_n g^n \;  ,
\ee
whose coefficients $b_n$ are, evidently, functions of the initial 
coefficients $a_n$ from series (2). Since the parameters of $f^*_k(g)$, 
according to Eq. (6), are defined through the re-expansion procedure,
the correcting function (10) contains not all powers of $g$ but only 
those starting from the order $k+1$, that is,
\be
\label{11} 
 C_{k+p}(g) = 1 + \sum_{n=k+1}^{k+p} b_n g^n \;  .
\ee
This fact essentially simplifies the construction of a self-similar 
approximant $C^*_{k+p}(g)$ from the correcting function (11). 
Constructing the approximant $C^*_{k+p}(g)$, we impose the limiting 
condition
\be
\label{12}
 \lim_{g\ra\infty} C_{k+p}^*(g) = const \; .
\ee

The corrected self-similar approximant is defined as 
\be
\label{13}
 \widetilde f_{k+p}(g) \equiv f_k^*(g) C_{k+p}^*(g) \; .
\ee
Because of condition (12),
\be
\label{14}   
 \lim_{g\ra\infty} \;
\frac{\ln C_{k+p}^*(g)}{\ln g} = 0 \; .
\ee
Therefore the strong-coupling exponent does not change,
\be
\label{15}
\lim_{g\ra\infty} \; \frac{\ln \widetilde f_{k+p}(g)}{\ln g} =
\lim_{g\ra\infty} \; \frac{\ln f_k^*(g)}{\ln g}  = \bt \; .
\ee
But the {\it corrected strong-coupling amplitude}
\be
\label{16}
\widetilde B_{k+p} \equiv \lim_{g\ra\infty} \; 
\frac{\widetilde f_{k+p}(g)}{ g^\bt}
\ee
changes to
\be
\label{17}
 \widetilde B_{k+p} = B_k \lim_{g\ra\infty}  C_{k+p}^*(g) \; ,
\ee
where $B_k$ is given by Eq. (8).

\subsection{Iterated root approximants}

In addition to self-similar exponential approximants [16,17] and factor 
approximants [18-21], there are self-similar root approximants [10-12].
The latter serve as a convenient tool for interpolating functions whose
strong-coupling behavior is described by means of known strong-coupling 
expansions, when the root approximants can be uniquely determined [22]. 
However, the parameters of the root approximants cannot be uniquely 
defined through the re-expansion procedure in the region of weak coupling.
Here we suggest a method allowing us to derive the parameters of the 
root approximants by means of the re-expansion procedure (6). 

The idea of the method, allowing for a unique derivation of the 
parameters of the root approximants, is as follows. In order to remove 
the multiplicity of solutions, arising in the re-expansion procedure  
applied to the general form of the root approximants, it is necessary to 
impose some restrictions on the definition of the related parameters. 
Thus, we can keep the same parameters of the lower-order approximants in 
the higher-order approximants, leaving there unknown only the 
highest-order parameter that is to be found from the re-expansion 
procedure. And the highest-order power has to be such that to satisfy 
the limiting condition (7). Since in this method, the lower-order root 
approximants are inserted into the higher-order approximants, the resulting
expressions can be called, {\it iterated root approximants}.       

To be concrete, let us assume that the factor $f_0(g)$ possesses the 
strong-coupling behavior as
\be
\label{18}
f_0(g) \simeq A g^\al \qquad ( g \ra \infty) \;  .
\ee
The first-order root approximant
\be
\label{19}
 R_1(g) = f_0(g) ( 1 + A_1 g)^\gm \; ,
\ee
where we set
\be
\label{20}
 \gm = \bt - \al \; ,
\ee
is the same as the first-order factor approximant and it is uniquely 
defined with $A_1 = a_1/\gamma$. In the second-order root approximant
\be
\label{21}
R_2(g) = f_0(g) \left ( \left ( 1 + A_1 g\right )^2 + A_2 g^2 
\right )^{\gm/2} \;  ,
\ee
we keep the same $A_1$ as before, and $\gamma$ is given by Eq. (20). 
Hence, we need to find only $A_2$ from the re-expansion procedure. 
In the third-order root approximant
\be
\label{22}
 R_3(g) = f_0(g) \left ( \left ( \left ( 1+ A_1 g \right )^2 +
A_2 g^2 \right )^{3/2} + A_3 g^3 \right )^{\gm/3} \; ,
\ee
the parameters $A_1$ and $A_2$ are kept the same as in the previous 
expression (21), while $A_3$ is defined through the re-expansion procedure.
Such an iteration construction continues, with the general $k$-approximant
being 
\be
\label{23}
 R_k(g) = f_0(g) \left ( \left ( \ldots \left ( 1 + A_1 g \right )^2 +
A_2 g^2 \right )^{3/2} + \ldots + A_k g^k \right )^{\gm/k} \;  ,
\ee
where all parameters $A_n$, with $n = 1, 2, \ldots, k-1$ are the same as in
the $k-1$-order approximant and the parameter $A_k$ is defined by the 
re-expansion procedure. 

Constructing in this way the iterated root approximants yields the 
strong-coupling amplitudes 
\be
\label{24}
 B_k = \lim_{g\ra\infty} \; \frac{R_k(g)}{g^\bt} \; .
\ee
In particular,
$$
B_1 = AA_1^\gm \; , \qquad B_2 = A \left ( A_1^2 + A_2\right )^{\gm/2} \; ,
\qquad
B_3 = A \left ( \left ( A_1^2 + A_2 \right )^{3/2} + A_3 
\right )^{\gm/3} \; ,
$$
and so on, with the $k$-order amplitude
\be
\label{25}
 B_k = A \left ( \left ( \ldots \left ( A_1^2 + A_2 \right )^{3/2}
+ A_3 \right )^{4/3} + \ldots + A_k \right )^{\gm/k} \; .
\ee
An accurate evaluation of the strong-coupling amplitude is the main aim of 
the suggested scheme. In the following sections, we shall consider several 
examples of explicit calculations of strong-coupling amplitudes, comparing
the accuracy of different approaches.

\section{Debye-H\"uckel function}

Let us start with the case, when the sought function is actually known. 
This will allow us to easily determine the accuracy of each scheme. Let 
us consider the Debye-H\"uckel function that is met in the theory of 
strong electrolites [23,24]. In dimensionless units, the function reads as
\be
\label{26}
 f(g) = \frac{2}{g} \; - \; \frac{2}{g^2} \left ( 1 - e^{-g}
\right ) \; .
\ee
In view of Eqs. (2)-(5), we have $f_0(g) = 1,\;\beta = -1$, and 
\be
\label{27}
 B = 2 \; .
\ee
The coefficients, corresponding to the weak-coupling expansion (2), are
$$
a_1 = - \; \frac{1}{3} \; , \qquad a_2 = \frac{1}{12} \; , \qquad 
a_3 = - \; \frac{1}{60} \; ,
$$
$$
a_4 = \frac{1}{360} \; , \qquad a_5 = -\; \frac{1}{2520} \; , \qquad 
a_6 = \frac{1}{20160} \; , \qquad a_7 = -\; \frac{1}{181440} \; .
$$
The Pad\'e approximants do not provide good accuracy. The best of 
them,
$$
P_{3/4}(g) = 
\frac{1+c_1g+c_2g^2+c_3 g^3}{1+c_4g+c_5g^2+C_6g^3+c_7g^4} \;  ,
$$
possessing the correct strong-coupling exponent $\beta = -1$, and invoking 
all seven expansion coefficients $a_n$, gives the strong-coupling amplitude 
$$
B_{3/4} = \lim_{g\ra\infty} g P_{3/4}(g) = \frac{c_3}{c_7} =
1.6 \;  ,
$$
with an error $\ep(B_{3/4}) = -20 \%$.

\subsection{Self-similar factor approximants}

The factor approximant, using the expansion terms up to the third order, that 
is, involving the coefficients $a_1, a_2, a_3$, reads as
\be
\label{28}
 f_3^*(g) = \left ( 1 + A_1 g\right )^{n_1}
\left ( 1 + A_2 g \right )^{n_2} \; ,
\ee
with the condition
\be
\label{29}
 n_1+n_2 = -1 \; .
\ee
This gives the strong-coupling amplitude
\be
\label{30}
 B_3 = A_1^{n_1} A_2^{n_2} = 1.64 \; ,
\ee
whose error is $\ep(B_3) = - 18 \%$.

The fifth-order factor approximant
\be
\label{31}
 f_5^*(g) = ( 1 + A_1 g)^{n_1}
(1 + A_2 g )^{n_2} (1 + A_3 g)^{n_3}  \; ,
\ee
invoking the weak-coupling expansion terms up to the fifth order in $g$ 
and satisfying the condition
\be
\label{32}
 n_1 + n_2 + n_3 = -1 \; ,
\ee
yields the strong-coupling amplitude
\be
\label{33}
 B_5 = A_1^{n_1} A_2^{n_2} A_3^{n_3} = 2.295\; ,
\ee
with an error $\ep(B_5)=15\%$.   

Note that the parameters of each of the factor approximants are defined 
through the re-expansion procedure. So, the values of $A_i$ and $n_i$ in 
each order are different. We use the same letters in different approximants
just for the simplicity of notation.

The factor approximant, using all seven expansion terms, is
\be
\label{34}
 f_7^*(g) = ( 1 + A_1 g)^{n_1}
(1 + A_2 g )^{n_2} (1 + A_3 g)^{n_3} (1+A_4 g)^{n_4} \;  ,
\ee
under the condition
\be
\label{35}
n_1 + n_2 + n_3 + n_4 = -1 \;   .
\ee
This gives the strong-coupling amplitude
\be
\label{36}
B_7 = A_1^{n_1} A_2^{n_2} A_3^{n_3} A_4^{n_4} = 1.799 \; ,
\ee
with an error $\ep(B_7)=-10\%$.

\subsection{Corrected factor approximants}

Following the scheme of Sec. 2, we can define the corrected factor 
approximants. To this end, let us introduce the correction function
\be
\label{37}
C_7(g) \simeq \frac{f_7(g)}{f_3^*(g)} \qquad (g\ra 0) \; ,
\ee
which yields
\be
\label{38}
 C_7(g) = 1 + b_4g^4 + b_5 g^5 + b_6g^6 + b_7 g^7 \; .
\ee
>From this expansion, we construct the factor approximant
\be
\label{39}
 C_7^*(g) = 1 + b_4g^4( 1+D_3g)^{n_3} (1+ D_4 g)^{n_4} \; ,
\ee
under the condition
\be
\label{40}
4 + n_3 + n_4 =0 \; .
\ee
The corrected factor approximant is defined as
\be
\label{41}
 \widetilde f_7(g) = f_3^*(g) C_7^*(g) \; .
\ee
This yields the strong-coupling amplitude
\be
\label{42}
 \widetilde B_7 = B_3 \left ( 1 + b_4 D_3^{n_3}D_4^{n_4} 
\right ) = 1.944 \; ,
\ee
with an error $\ep(\widetilde{B}_7) = - 2.8 \%$.

\subsection{Simple root approximants}

A simple root approximant
\be
\label{43}
r_3^*(g) = \left ( \left ( 1 + A_1 g \right )^{n_1} +
A_2 g^2 \right )^{-1/2} \; ,
\ee
in which the parameters are defined trough the re-expansion procedure,
gives the strong-coupling amplitude
\be
\label{44}
 B_3 = \frac{1}{\sqrt{A_2}} = 2.164 \; ,
\ee
whose error is $\ep(B_3) = 8.2 \%$.

\subsection{Corrected root approximants}

The above simple root approximant can be corrected according to Sec. 2.
For this purpose, we define the correction function
\be
\label{45}
 C_7(g) \simeq \frac{f_7(g)}{r_3^*(g)} \qquad (g\ra 0) \; ,
\ee
corresponding to the expansion in powers of $g$ up to the seventh order,
and construct the root approximant
\be
\label{46}
C_7^*(g) = 1 + b_4 g^4 \left ( ( 1+D_3 g)^{n_3} +
D_4 g^2 \right )^{-2} \;  .
\ee
Then the corrected root approximant is
\be
\label{47}
 r_7^*(g) = r_3^*(g) C_7^*(g) \; .
\ee
The latter gives the strong-coupling amplitude
\be
\label{48}
 \widetilde B_7 = \frac{1}{\sqrt{A_2} } \left ( 1 +
\frac{b_4}{D_4^2} \right ) = 1.875 \; ,
\ee
with an error $\ep(\widetilde{B}_7) = - 6.3 \%$.

\subsection{Iterated root approximants}

The iterated root approximants, according to Sec. 2, have the form
\be
\label{49}
R_k(g) = \left ( \left ( \ldots ( 1 + A_1 g)^2 + A_2 g^2 
\right )^{3/2} + \ldots + A_k g^k \right )^{\bt/k} \; ,   
\ee
where $\beta = -1$. Therefore, the strong-coupling amplitude of $k$-order is
\be
\label{50}
B_k = \left ( \left ( \ldots  \left ( A_1^2 + A_2 \right )^{3/2} +
A_3 \right )^{4/3} + \ldots + A_k \right )^{\bt/k} \;  .
\ee
>From here, we find
$$
B_1 = 3 \; , \qquad B_2 = 2.449 \; , \qquad B_3 = 2.229 \; , \qquad
B_4 = 2.127 \; ,
$$
$$
B_5 = 2.067 \; , \qquad B_6 = 2.032 \; , \qquad B_7 = 2.009 \;   .
$$
The seventh order amplitude has an error $\ep(B_7) = 0.45 \%$.
Thus, the iterated root approximant (49) of the seven-th order is the most 
accurate among all above considered approximants.

\section{Quartic anharmonic oscillator}

The anharmonic oscillator, with the Hamiltonian
\be
\label{51}
H = - \; \frac{1}{2} \; \frac{d^2}{dx^2} + \frac{1}{2} \; x^2 +
gx^4 \;  ,
\ee
is the model imitating many systems in quantum chemistry, atomic physics, 
condensed-matter physics, and field theory. Here, $g$ is the coupling 
parameter and we set $x \in (-\infty, \infty)$ and $g \in [0, \infty)$. 
The ground-state energy for this Hamiltonian, found by means of the 
weak-coupling perturbation theory [25,26], when $g \rightarrow 0$, reads as
\be
\label{52}
 e(g) \simeq a_0 + a_1 g + a_2 g^2 + a_3 g^3 +
a_4 g^4 + a_5 g^5 + a_6 g^6 + a_7 g^7 \; ,
\ee
with the coefficients
$$
 a_0 =\frac{1}{2} \; , \qquad a_1 = \frac{3}{4} \; , \qquad
a_2 = -\; \frac{21}{8} \; , \qquad a_3 = \frac{333}{16} \; ,
$$
$$
a_4 = - \; \frac{30885}{128} \; , \qquad a_5 = \frac{916731}{256} \; ,
\qquad a_6 = -\; \frac{65518401}{1024} \; , \qquad
a_7 = \frac{2723294673}{2048} \; .
$$
The strong-coupling limit for the ground-state energy yields
\be
\label{53}
 e(g) \simeq 0.667986\; g^{1/3} \qquad (g \ra \infty) \; .
\ee
Hence, the strong-coupling exponent $\beta = 1/3$ and the strong-coupling 
amplitude is 
\be
\label{54}
B = 0.667986 \;  .
\ee

As is seen, the coefficients of the weak-coupling expansion quickly 
grow, and perturbation theory is divergent for any nonzero $g$. The 
standard Pad\'e approximants for this problem are not applicable 
at all, being unable to satisfy the fractional strong-coupling exponent 
$\beta =1/3$. It is possible to introduce the power-transformed 
approximants, similarly to Ref. [14], fixing the exponent $\beta =1/3$. 
But this way does not provide a great accuracy improvement. The best 
power-transformed Pad\'e approximant gives an error of $8\%$.

\subsection{Self-similar factor approximants}

Using the third order of the weak-coupling expansion (52), we get the
self-similar factor approximant
\be
\label{55}
e_3^*(g) = \frac{1}{2} (1 + A_1 g)^{n_1} ( 1+ A_2 g)^{n_2} \;  ,
\ee
under the condition
\be
\label{56}
 n_1 + n_2 = \frac{1}{3} \; .
\ee
This gives the strong-coupling amplitude
\be
\label{57}
 B_3 = \frac{1}{2} \; A_1^{n_1} A_2^{n_2} = 0.75 \; ,
\ee
with an error $\ep(B_3) = 12 \%$. 

The fifth-order factor approximant has the form
\be
\label{58}
 e_5^*(g) = \frac{1}{2} (1 + A_1 g)^{n_1} ( 1+ A_2 g)^{n_2} 
(1 + A_3 g)^{n_3}\;  ,
\ee
with the condition
\be
\label{59}
 n_1 + n_2 + n_3 = \frac{1}{3} \;   .
\ee
Then the strong-coupling amplitude is
\be
\label{60}
 B_5 = \frac{1}{2} \; A_1^{n_1} A_2^{n_2} A_3^{n_3} = 0.725 \; ,
\ee
whose error is $\ep(B_5) = 8.5 \%$.

The seventh-order factor approximant is
\be
\label{61}
  e_7^*(g) = \frac{1}{2} (1 + A_1 g)^{n_1} ( 1+ A_2 g)^{n_2} 
(1 + A_3 g)^{n_3} (1+ A_4 g)^{n_4} \; ,
\ee
with the condition
\be
\label{62}
n_1 + n_2 + n_3 + n_4 = \frac{1}{3} \; .
\ee
>From here, we have the strong-coupling amplitude
\be
\label{63}
B_7 = \frac{1}{2} \; A_1^{n_1} A_2^{n_2} A_3^{n_3} A_4^{n_4}
= 0.712 \;   ,
\ee
with an error $\ep(B_7) = 6.6 \%$.

\subsection{Corrected factor approximants}

Defining the correction function
\be
\label{64}
C_7(g) \simeq \frac{e_7(g)}{e_3^*(g)} \qquad (g\ra 0) \;  ,
\ee
we get the series
\be
\label{65}
C_7(g) = 1 + \sum_{n=4}^7 b_n g^n \;  .
\ee
The latter generates the factor approximant
\be
\label{66}
C_7^*(g) = 1 +
 b_4 g^4 ( 1 + D_3 g)^{n_3} ( 1 + D_4 g)^{n_4} \;  ,
\ee
where
\be
\label{67}
4 + n_3 + n_4 = 0 \;  .
\ee
 
The corrected factor approximant
\be
\label{68}
\widetilde e_7(g) = e_3^*(g) C_7^*(g)
\ee
yields the strong-coupling amplitude
\be
\label{69}
\widetilde B_7 = \frac{1}{2}\;
A_1^{n_1} A_2^{n_2} \left ( 1 + b_4 D_3^{n_3} D_4^{n_4}
\right ) = 0.728 \;  ,
\ee
with an error $\ep(\widetilde{B}_7)=9\%)$.

\subsection{Simple root approximants}

Simple root approximants do not give good accuracy. For instance, 
the root approximant
\be
\label{70}
r_5^*(g) = \frac{1}{2} \left ( \left ( \left ( 1 + A_1 g 
\right )^{n_1} + A_2 g^2 \right )^2 + A_3 g^3 \right )^{1/9}
\ee
results if the strong-coupling amplitude
\be
\label{71}
 B_5 = \frac{1}{2} \; A_3^{1/9} = 0.824 \; ,
\ee
whose accuracy is characterized by the error $\ep(B_5) = 23\%$.

\subsection{Iterated root approximants}

According to Sec. 2, the iterated root approximants are given by  
$$
R_1(g) = \frac{1}{2} ( 1 + A_1 g )^{1/3} \; , \qquad
R_2(g) = \frac{1}{2} \left ( ( 1 + A_1 g )^2 + A_2 g^2 
\right )^{1/6} \; ,
$$
\be
\label{72}
 R_3(g) = \frac{1}{2} \left ( \left (  ( 1 + A_1 g )^2 + 
A_2 g^2 \right )^{3/2} + A_3^3 \right )^{1/9} \;  ,   
\ee
and so on. However, the approximant $R_4(g)$ is complex, and we limit 
ourselves by Eqs. (72). The related strong-coupling amplitudes are
$$
B_1 = \frac{1}{2} \; A_1^{1/3} = 0.825 \; , \qquad
B_2 = \frac{1}{2} \left ( A_1^2 + A_2 \right )^{1/6} = 0.572 \; ,
$$
\be
\label{73}
B_3 = \frac{1}{2} \left ( \left ( A_1^2 + A_2 \right )^{3/2}
+ A_3 \right )^{1/9} = 0.855 \;  .
\ee
Their errors are $\ep(B_1)=24\%, \;\ep(B_2)=-14\%,\; \ep(B_3)=28\%$.

\subsection{Power-transformed approximants}

It is possible to define the power transformation [14] as
\be
\label{74}
P_k(g,m) \simeq f_k^m(g) \qquad (g \ra 0) \;  ,
\ee
producing the series
$$
P_k(g,m) = \sum_{n=0}^k b_n(m) g^n \;  .
$$
Then, constructing a self-similar approximation on the basis of this 
series, we get $P_k^*(g,m)$. The final answer is given by the inverse 
power transformation 
\be
\label{75}
 f_k^*(g) = \left [ P_k^*(g,m) \right ]^{1/m} \; .
\ee
This way was analized in Ref. [14], where the transformation power $m$
was defined by variational optimization. Such a way, however, is rather 
cumbersome and yields the accuracy improvement not better than the 
simpler methods considered here.

Among the methods, studied in this Section, the self-similar factor 
approximants provide the best accuracy for the strong-coupling amplitude.

\section{Polymer expansion factor}

The expansion factor $\alpha(g)$, as a function of the coupling parameter 
$g$, for a polymer chain with excluded interactions, can be calculated by 
means of perturbation theory [27] resulting in the series
\be
\label{76}
 \al(g) = 1 + a_1 g + a_2 g^2 + a_3 g^3 + a_4 g^4 +
+ a_5 g^5 + a_6 g^6 \; ,
\ee
with the coefficients
$$
a_1 = \frac{4}{3} \; , \qquad a_2 = -2.075385396 \; ,
\qquad a_3 = 6.296879676 \; ,
$$
$$ 
a_4 = -25.05725072 \; , \qquad a_5 = 116.134785 \;,
\qquad  a_6 = -594.71663 \; .
$$
Numerical fitting [27,28] gives the phenomenological formula
\be
\label{77}
 \al(g) = \left ( 1 + 7.524 g + 11.06 g^2 \right )^{0.1772} \; .
\ee
This implies that the strong-coupling exponent $\beta = 0.3544$, which 
is in agreement with other numerical simulations [29], where the exponent
\be
\label{78}
\nu \equiv \frac{1}{2} \left ( 1 + \frac{\bt}{2} \right )
\ee
was calculated, giving $\nu = 0.5877$. Therefore the strong-coupling 
amplitude is
\be
\label{79}
 B = 1.531 \; .
\ee

\subsection{Self-similar factor approximants}

The self-similar factor approximants start with the lowest order
\be
\label{80}
 \al_1^*(g) = (1 + A_1 g)^\bt \; ,
\ee
in which $A_1 = a_1/\beta$ and $\beta = 0.3544$. Then the strong-coupling 
amplitude is
\be
\label{81}
B_1 = A_1^\bt = 1.599 \;  ,
\ee
with an error $\ep(B_1) = 4.4 \%$. Increasing the approximation order
improves the accuracy. Thus, the fifth-order approximant
\be
\label{82}
\al_5^*(g) = (1 + A_1 g)^{n_1} (1 + A_2 g)^{n_2}
(1 + A_3 g)^{n_3} \;  ,
\ee 
where
\be
\label{83}
 n_1 + n_2 + n_3 = \bt = 0.3544 \; ,
\ee
gives the strong-coupling amplitude
\be
\label{84}
B_5 = A_1^{n_1} A_2^{n_2} A_3^{n_3} = 1.541 \;  ,
\ee
with an error $\ep(B_5) = 0.65 \%$.

\subsection{Corrected factor approximants}

Introducing the correction function
\be
\label{85}
C_3(g) \simeq \frac{\al_3(g)}{\al_1^*(g)} \qquad 
(g \ra 0) \;  ,
\ee
and constructing the related factor approximant
\be
\label{86}
C_3^*(g) = 1 + b_2 g^2 ( 1 + D_1 g)^{-2} \;  ,
\ee
we come to the corrected factor approximant
\be
\label{87}
\widetilde\al_3(g) = \al_1^*(g) C_3^*(g) \;  .
\ee
This gives the strong-coupling amplitude
\be
\label{88}
\widetilde B_3 = A_1 ^\bt \left ( 1 +
\frac{b_2}{D_1^2} \right ) = 1.552 \;  ,
\ee
with an error $\ep(\widetilde{B}_3) = 1.4 \%$.

The accuracy of the corrected factor approximants is close to that of 
the directly constructed self-similar factor approximants.

\subsection{Simple root approximants}

Simple root approximants provide sufficiently good accuracy. Thus the
approximant
\be
\label{89}
r_5^*(g) = \left ( \left ( ( 1 + A_1 g)^{n_1} + A_2 g^2 
\right )^{n_2} + A_3 g^3 \right )^{\bt/3}
\ee
yields the strong-coupling amplitude
\be
\label{90}
 B_5 = A_3^{\bt/3} = 1.537 \; ,
\ee
with an error $\ep(B_5) = 0.4 \%$.

\subsection{Corrected root approximants}

For the correction function
\be
\label{91}
C_5(g) \simeq \frac{\al_5(g)}{\al_1^*(g)} \qquad 
( g \ra 0) \;  ,
\ee
the corresponding self-similar root approximant reads as
\be
\label{92}
 C_5^*(g) = 1 + b_2 g^2 \left ( ( 1 + D_1 g)^{m_1} + 
D_2 g^2 \right )^{-1} \; .
\ee
The corrected root approximant
\be
\label{93}
\widetilde r_5(g) = \al_1^*(g) C_5^*(g)   
\ee
results in the strong-coupling amplitude
\be
\label{94}
 \widetilde B_5 = A_1^\bt \left ( 1 + \frac{b_2}{D_2}
\right ) = 1.542 \; ,
\ee
whose error is $\ep(\widetilde{B}_5) = 0.72 \%$.

The simple root approximants, being already sufficiently accurate, are not 
improved by the corrected variants.

\subsection{Iterated root approximants}

For the iterated root approximants, following the scheme of Sec. 2, we find
the strong-coupling amplitudes
$$
B_1 = 1.599 \; , \qquad B_2 = 1.544 \; , \qquad B_3 = 1.549 \; ,
$$
\be
\label{95}
B_4 = 1.539 \; , \qquad B_5 = 1.541 \; , \qquad B_6 = 1.537 \;   .
\ee
The error of the highest-order approximant is $\ep(B_6) = 0.4 \%$.

All variants of the approximants, studied in the present Section, are of 
about the same accuracy. We have also considered the power-transformed 
approximants [14], whose accuracy is found to be close to that of the 
iterated root approximants.

\section{Oscillating fluid string}

In biological and chemical applications, there exists an important class 
of membranes called fluid membranes [30]. The model of a one-dimensional 
fluid string, oscillating between two walls, was advanced by Edwards [31]. 
It has been shown [31-33] that the free energy of the string coincides 
with the ground-state energy of a quantum particle in a one-dimensional 
box. This energy, as a function of a finite wall stiffness $g$, reads as
\be
\label{96}
E(g) = \frac{\pi^2}{8g^2} \left ( 1 + \frac{g^2}{32} +
\frac{g}{4} \; \sqrt{ 1 + \frac{g^2}{64} } \right ) \; .
\ee
In the small-stiffness expansion
\be
\label{97}
 E_k(g) =  \frac{\pi^2}{8g^2}\; \sum_{n=0}^k a_n g^n \; ,
\ee
the coefficients are
$$
a_0 = 1 \; , \qquad a_1 = \frac{1}{4}\; , \qquad 
a_2 = \frac{1}{32} \; ,\qquad a_3 = \frac{1}{512} \; ,
$$
$$
a_4 = 0 \; , \qquad a_5 =-\; \frac{1}{131072} \; , 
\qquad a_6 = 0 \; , \qquad a_7 = \frac{1}{16777216} \;  .
$$
The rigid walls correspond to the stiffness $g\ra\infty$, when
\be
\label{98}
E(\infty) = \frac{\pi^2}{128} = B = 0.077106 \; .
\ee
Hence, the strong-stiffness exponent is $\beta = 0$.

Our aim is to approximate the rigid-wall limit (98) employing the 
small-stiffness expansion (97).

Pad\'e approximants are not applicable for this problem, giving 
negative values of the strong-stiffness amplitude. So, we shall apply 
the self-similar approximants.

\subsection{Simple root approximants}

The root approximants of low orders are not sufficiently accurate. 
Thus, the root approximant
\be
\label{99}
 r_3^*(g) = E_0(g) \left ( ( 1 + A_1 g )^{n_1} + A_2 g^2 
\right ) \;  ,
\ee
where $E_0(g) = \pi^2/8g^2$, gives the strong-stiffness amplitude
\be
\label{100}
B_3 = \frac{\pi^2}{8} \; A_2 = 0.0544 \;  ,
\ee
with an error $\ep(B_3) = -29\%$. And the higher-order root 
approximants are not convenient, being not uniquely defined through 
the re-expansion procedure.

\subsection{Corrected root approximants}

As is shown in Sec. 2, we can introduce the corrected root approximants.
For instance, the correction function
\be
\label{101}
C_7(g) \simeq \frac{E_7(g)}{r_3^*(g)} \qquad (g \ra 0 )
\ee
generates the self-similar root approximant
\be
\label{102}
 C_7^*(g) = 1 + b_4 g^4 \left ( ( 1 + D_1 g)^{m_1} + 
D_2 g^2 \right )^{-2} \;  .
\ee
Then the corrected root approximant
\be
\label{103}
r_7^*(g) = r_3^*(g) C_7^*(g)
\ee
yields the strong-stiffness amplitude 
\be
\label{104}
\widetilde B_7 = \frac{\pi^2}{8} \; A_2 \left ( 1 +
\frac{b_4}{D_2^2} \right ) = 0.0771 \; ,
\ee
with an error $\ep(\widetilde{B}_7)=-0.02\%$.

We may note that the sixth-order corrected root approximant is also 
of good accuracy, with an error $\widetilde{B}_6=-0.66 \%$.

\subsection{Iterated root approximants}

The uniquely defined iterated root approximants, introduced in Sec. 2,  
result in the strong-stiffness amplitudes
$$
B_1 = 0.019 \; , \qquad B_2 = 0.039 \; , \qquad B_3 = 0.051 \; ,
$$
\be
\label{105}
B_4 = 0.058 \; , \qquad B_5 = 0.062 \; , \qquad B_6 = 0.065 \; ,
\qquad B_7 = 0.067 \;  .
\ee
The errors of the highest two approximations are $\ep(B_6) = -16 \%$
and $\ep(B_7) = -13 \%$, respectively.

Another variant was considered in Ref. [15], invoking the change of 
the variable $g$, and giving the strong-stiffness amplitude of seventh 
order with an error of $5 \%$. The best accuracy is provided by the 
corrected root approximant (103) yielding amplitude (104).

\section{Fluctuating fluid membrane}

For a three-dimensional fluctuating fluid membrane, there are no exact 
results. Thus, the pressure of this membrane is calculated by means of 
perturbation theory [34] yielding expansions 
\be
\label{106}
p_k(g) = \frac{\pi^2}{8g^2} \; \sum_{n=0}^k a_n g^n \; .
\ee
The coefficients are known [34] up to the sixth order:
$$
a_0 = 1 \; , \qquad a_1 = \frac{1}{4} \; , \qquad 
a_2 = \frac{1}{32} \; , \qquad a_3 = 2.176347\times 10^{-3} \; , 
$$
$$\qquad a_4 = 0.552721 \times 10^{-4} \; , \qquad
a_5 = - 0.721482 \times 10^{-5} \; , \qquad
a_6 = - 1.777 848 \times 10^{-6} \;  .
$$
The strong-stiffness limit has been found by Monte Carlo simulations [35],
giving
\be
\label{107}
 p(\infty) = 0.0798 \pm 0.0003 \; .
\ee
This implies that the strong-stiffness exponent is $\beta = 0$ and the 
strong-stiffness amplitude is
\be
\label{108}
B = 0.0798 \;  .
\ee

Pad\'e approximants are inapplicable for this problem, resulting 
in negative values for $B$.

\subsection{Simple root approximants}

The simple root approximant
\be
\label{109}
r_3^*(g) = \frac{\pi^2}{8g^2} \left ( ( 1 + A_1 g)^{n_1}
+ A_2 g^2 \right )
\ee
gives the strong-stiffness amplitude
\be
\label{110}
 B_3 = \frac{\pi^2}{8} \; A_2 = 0.056 \; ,
\ee
with an error $\ep(B_3) = -30 \%$. The higher-order root 
approximants are not convenient to employ, since they are not uniquely 
defined.

\subsection{Corrected root approximants}

If we set $a_7 = 0$, we can introduce the correction function
\be
\label{111}
C_7(g) \simeq \frac{p_7(g)}{r_3^*(g)} \qquad (g\ra 0) \;  .
\ee
The corresponding self-similar root approximant is
\be
\label{112}
C_7^*(g) = 1 + b_4 g^4 \left ( ( 1+ D_1 g)^{m_1}
+ D_2 g^2 \right )^{-2} \; .
\ee
The corrected root approximant is defined as
\be
\label{113}
 \widetilde p_7(g) = r_3^*(g) C_7^*(g) \; ,
\ee
which gives the strong-stiffness amplitude
\be
\label{114}
 \widetilde B_7 = \frac{\pi^2}{8} \; A_2 \left (
1 + \frac{b_4}{D_2^2} \right ) = 0.085 \; ,  
\ee
with an error $\ep(\widetilde{B}_7) = 6.5\%$.

\subsection{Iterated root approximants}

Employing the iterated root approximants of Sec. 2, we have
$$
B_1 = 0.019 \; , \qquad B_2 = 0.039 \; , \qquad 
B_3 = 0.053 \; ,
$$
\be
\label{115}
B_4 = 0.061 \; , \qquad B_5 = 0.067 \; , \qquad 
B_6 = 0.071 \;   .
\ee
The error of $B_6$ is $\ep(B_6) = -11 \%$.

In this way, for this problem, the corrected root approximants are 
the most accurate ones.

\section{Strongly interacting fermions}

The ground-state energy of a diluted Fermi gas, using perturbation theory,
can be represented [36,37] as an asymptotic expansion 
\be
\label{116}
e(g) \simeq a_0 + a_1 g + a_2 g^2 + a_3 g^3 + a_4 g^4
\ee
in powers of the dimensionless coupling parameter
\be
\label{117}
 g \equiv | k_F a_s | \;  ,
\ee
in which $k_F$ is a Fermi wave vector and $a_s$, scattering length.
Here $g \rightarrow 0$ and the coefficients are
$$
a_0 = \frac{3}{10} \; , \qquad a_1 = - \; \frac{1}{3\pi} \; ,
\qquad a_2 = 0.055661 \; ,
$$
$$
a_3 = - 0.00914 \; , \qquad a_4 = - 0.018604 \; .
$$
Numerical calculations [38,39] give the strong-coupling limit
\be
\label{118}
 \lim_{g\ra\infty} e(g) = 0.132 \; .
\ee
Hence, the strong-coupling exponent is $\beta = 0$, while the 
strong-coupling amplitude is
\be
\label{119}
 B =0.132 \; .
\ee

Pad\'e approximants are not accurate for this problem, the best 
of them having an error of $30 \%$.

\subsection{Self-similar factor approximants}

The third-order factor approximant reads as
\be
\label{120}
 e_3^*(g) = \frac{3}{10} ( 1 + A_1 g)^{n_1}
( 1 + A_2 g )^{n_2} \;  ,
\ee
under the condition
\be
\label{121}
n_1 = n_2 = 0 \;  .
\ee
This yields the strong-coupling amplitude
\be
\label{122}
B_3 = \frac{3}{10} \; A_1^{n_1} A_2^{n_2} = 0.174 \;  ,
\ee
with an error $\ep(B_3) = 32 \%$.

The fourth-order factor approximant can be written as 
\be
\label{123}
 e_4^*(g) = \frac{3}{4} + a_1 g (1 + A_3 g)^{n_3}
(1 + A_4 g)^{n_4} \; ,
\ee
under the condition
\be
\label{124}
1 + n_3 + n_4 = 0 \;  .
\ee
Then the strong-coupling amplitude becomes
\be
\label{125}
 B_4 = \frac{3}{10} + a_1 A_3^{n_3} A_4^{n_4} =
0.162 \; ,
\ee
with an error $\ep(B_4)=23\%$.

\subsection{Corrected factor approximants}

For the correction function
\be
\label{126}
 C_4(g) \simeq \frac{e_4(g)}{e_3^*(g)} \qquad
(g\ra 0) \; ,
\ee
the factor approximant can be represented as
\be
\label{127}
 C_4^*(g) = 1 + b_4 g^4 ( 1 + D_1 g)^{-4} \; .
\ee
The corrected factor approximant
\be
\label{128}
\widetilde e_4(g) = e_3^*(g) C_4^*(g)
\ee
yields the strong-coupling amplitude
\be
\label{129}
\widetilde B_4 = \frac{3}{10} \; A_1^{n_1} A_2^{n_2} \left (
1 + \frac{b_1}{D_1^4} \right ) = 0.1434 \; ,
\ee
whose error is $\ep(\widetilde{B}_4)=8.6\%$.

\subsection{Iterated root approximants}

The iterated root approximants, as defined in Sec. 2, exist for the 
lowest two approximations, giving
\be
\label{130}
 B_1 = 0.098 \;, \qquad B_2=0.169 \; .
\ee
The higher approximants become complex. The accuracy of $B_2$ is 
characterized by an error $\ep(B_2) = 28 \%$.

In this way, the best approximation is provided by the corrected 
factor approximant (128), with an error of $8.6 \%$.

\section{One-dimensional Bose system}

The one-dimensional Bose gas with local interactions corresponds to
the Lieb-Liniger model [40]. The ground-state energy of the gas, in 
the weak-coupling limit, is described by the expansion
\be
\label{131}
 e(g) \simeq g + a_3 g^{3/2} + a_4 g^2 + a_5 g^{5/2} \; ,
\ee
with the coefficients
$$
 a_3 = - \; \frac{4}{3\pi} = 0.424413 \; , \qquad
 a_4 = \frac{1.29}{2\pi^2} = 0.065352 \; , \qquad
 a_5 = - 0.017201 \;.
$$
In the strong-coupling limit, one gets the Tonks-Girardeau expression
\be
\label{132}
 \lim_{g\ra\infty} e(g) = \frac{\pi^2}{3} = B = 3.289868 \; .
\ee
More details can be found in the review [41].

\subsection{Simple root approximants}

The simple root approximant
\be
\label{133}
r_3^*(g) = g \left ( \left ( 1 + A_1 g^{1/2} \right )^{n_1}
+ A_2 g \right )^{-1}
\ee
gives the strong-coupling amplitude
\be
\label{134}
B_3 = \frac{1}{A_2} = 4.2835 \;  ,
\ee
with an error $\ep(B_3) = 48 \%$. Certainly, such an accuracy is 
not sufficient.

\subsection{Iterated root approximants}

Using the iterated root approximants of Sec. 2, with setting $a_6 = 0$,
we get the strong-coupling amplitudes
\be
\label{135}
 B_2 = 8.713 \; , \qquad B_3 = 4.765 \; , \qquad
B_4 = 3.2924 \; .
\ee
The error of the highest-order approximation is $\ep(B_4)=0.08\%$.
This is the best accuracy among the considered approximants.

\subsection{Inversion of strong-coupling series}

It is worth noting that in those cases when several terms of the 
strong-coupling expansion are known, it is possible to change the 
variable and to treat the old strong-coupling expansion as a 
weak-coupling one in terms of the new variable, and vice versa, 
the old weak-coupling expansion, as a strong-coupling one. For 
instance, the strong-coupling expansion for the ground-state energy 
of the Lieb-Liniger model can be written (see Ref. [41]) in the form
$$
 e(g) \simeq \frac{\pi^2}{3} \left (
1 \; - \; \frac{4}{g} \; + \; \frac{12}{g^2} \; - \;
\frac{32}{g^3} \; + \; \frac{80}{g^4} \right ) \; .
$$  
Changing the variable to $x=1/g$ makes the above series a weak-coupling
expansion in terms of $x$. Conversely, Eq. (131) transforms, in terms of 
$x$, into a strong-coupling expansion [15]. However, possessing quite a 
number of terms of the strong-coupling limit with respect to the coupling 
parameter $g$ is a rather rare case. Therefore, in the present paper, we 
focus on the problem of dealing with the direct coupling parameter, 
without changing the variable.

\section{Bose-Einstein condensation temperature}

An important problem is the dependence of the Bose-Einstein condensation 
temperature $T_c$ on the gas parameter
\be
\label{136}
 \gm \equiv \rho^{1/3} a_s \;  ,
\ee
in which $\rho$ is the average atomic density and $a_s$, scattering 
length. For the asymptotically weak interaction, when $\gm\ra 0$,
the relative shift of the critical temperature, with respect to the 
condensation temperature of the ideal Bose gas,
$$
T_0 = \frac{2\pi\hbar^2}{mk_B} \left [
\frac{\rho}{\zeta(3/2)} \right ]^{2/3} \;  ,
$$
is linear in the gas parameter:
\be
\label{137}
\frac{\Dlt T_c}{T_0} \simeq c_1 \gm \qquad (\gm\ra 0) \; ,
\ee
where
$$
\Dlt T_c \equiv T_c - T_0 \;  .
$$

The coefficient $c_1$ has been calculated by a number of different 
methods (see review articles [42,43]). It can be represented as an 
asymptotic expansion
\be
\label{138}
c_1(g) \simeq a_1 g + a_2 g^2 + a_3 g^3 + a_4 g^4 + a_5 g^5
\ee
in powers of an effective coupling parameter [44-46], with the 
coefficients
$$
a_1 = 0.223286 \; , \qquad a_2 =-0.0661032 \; , \qquad
a_3 = 0.026446 \; ,
$$
$$
a_4 = -0.0129177 \; , \qquad a_5 = 0.00729073 \;  .
$$  
The sought value of $c_1$ is the limit 
\be
\label{139}
c_1 = \lim_{g\ra\infty} c_1(g) = B \; .
\ee
   
The most accurate calculations of $c_1$ have been accomplished by means 
of the Monte Carlo simulations in Refs. [47-51] and, by using different 
variants of the optimized perturbation theory [52], in Refs. [44-46,53-55]. 
Summarizing  these results, we have:
$$
c_1 = 1.32 \pm 0.02 \; , \qquad [47,48]
$$
$$
c_2 = 1.29 \pm 0.05 \; , \qquad [49,50]
$$
$$
c_1 = 1.32 \pm 0.14 \; , \qquad [51]
$$
$$
c_1 = 1.27 \pm 0.11 \; , \qquad [45,46]
$$
$$
c_1 = 1.30 \pm 0.03 \; . \qquad [54,55]
$$

The use of Pad\'e approximants, for summing the asymptotic series 
(138), does not lead to accurate approximations, the best approximant 
having an error of $-24 \%$.

\subsection{Self-similar factor approximants}

The direct application of the self-similar factor approximants does not
provide high accuracy. Thus, the factor approximant
\be
\label{140}
 f_3^*(g) = a_1 g ( 1 + A_1 g)^{n_1} ( 1 + A_2 g)^{n_2} \; ,
\ee
under the condition
\be
\label{141}
1 + n_1 + n_2 = 0 \;  ,
\ee
gives the amplitude
\be
\label{142}
B_3 = a_1 A_1^{n_1} A_2^{n_2} = 1.025 \;  ,
\ee
which approximates limit (139) with an error $\ep(B_3)=-21\%$.

\subsection{Simple root approximants}

Simple root approximants also are not accurate enough. For example, 
the root approximant
\be
\label{143}
r_2^*(g) = a_1 g ( 1 + A_1 g)^{-1}
\ee
yields the amplitude
\be
\label{144}
B_2 = \frac{a_1}{A_1} = - \; \frac{a_1}{a_2} = 0.754 \; ,
\ee
with an error $\ep(B_2) = -42 \%$. And the root approximant
\be
\label{145}
r_3^*(g) = a_1 g \left ( ( 1 + A_1 g )^{n_1} + A_2 g^2 
\right )^{-1/2}
\ee
gives the amplitude
\be
\label{146}
B_3 = \frac{a_1}{\sqrt{A_2}} = 0.916 \; ,
\ee
whose accuracy is only sightly better, having an error 
$\ep(B_3) = -30 \%$.

\subsection{Iterated root approximants}

The iterated root approximants of Sec. 2 result in the amplitudes
\be
\label{147}
B_1 = 0.754 \; , \qquad B_2 = 1.383 \; , \qquad
B_3 = 0.854 \;  .
\ee
The last of these approximations has an error $\ep(B_3) = -34 \%$.
The iterated root approximant of the fourth order does not exist, 
being complex valued.

\subsection{Corrected root approximants}

For the third-order correction function
\be
\label{148}
 C_3(g) \simeq \frac{a_1g + a_2 g^2 + a_3 g^3}{r_2^*(g)}
\qquad ( g\ra 0) \; ,
\ee
the corresponding root approximant is
\be
\label{149}
 C_3^*(g) = 1 + b_2 g^2 ( 1 + D_1 g)^{-2} \; .
\ee
The corrected root approximant becomes
\be
\label{150}
\widetilde r_3(g) = r_2^*(g) C_3^*(g) \;  .
\ee
This gives the amplitude
\be
\label{151}
\widetilde B_3 = - \; \frac{a_1}{a_2} \left (
1 + \frac{b_1}{D_1^2} \right ) = 0.924 \; ,
\ee
with an error $\ep(\widetilde{B}_3) = -29 \%$.

Increasing the order of the correction function essentially improves 
the accuracy. Thus, for the fifth-order correction function
\be
\label{152}
C_5(g) \simeq 
\frac{a_1g+a_2g^2+a_3g^3+a_4 g^4+a_5 g^5}{r_2^*(g)} \; ,   
\ee
whose root approximant is
\be
\label{153}
 C_5^*(g) = 1 + b_2 g^2 \left ( ( 1 + D_1 g)^2 +
D_2 g^2 \right )^{-1}  \;   ,
\ee
we have the corrected root approximant
\be
\label{154}
\widetilde r_5(g) = r_2^*(g) C_5^*(g) \;  .
\ee
The latter yields the amplitude
\be
\label{155}
 \widetilde B_5 = -\; \frac{a_1}{a_2} \left (
1 + \frac{b_2}{D_1^2+D_2} \right ) = 1.29 \; ,
\ee
which is in the frame of the numerically calculated values of $c_1$.

\subsection{Corrected-iterated root approximants}

The iterated root approximants, considered in subsection 10.3, are 
not accurate enough. But it is possible to correct them in the spirit 
of Sec. 2. Let us take, for example, the third-order iterated root 
approximant
\be
\label{156}
R_3(g) = a_1 g \left ( ( 1 + A_1 g )^2 + 
A_2 g^2 \right )^{-1/2} \;  .
\ee
Introduce the correction function
\be
\label{157}
 C_5(g) \simeq 
\frac{a_1g+a_2g^2+a_3g^3+a_4 g^4+a_5 g^5}{R_3(g)} \;  ,
\ee 
whose self-similar approximant is
\be
\label{158}
  C_5^*(g) = 1 + b_3 g^3 ( 1 + D_3 g )^{-3} \; .
\ee
The corrected-iterated root approximant 
\be
\label{159}
R_5^*(g) = R_3(g) C_5^*(g)
\ee
gives the amplitude
\be
\label{160}
B_5^* = a_1 \left ( A_1^2 + A_2 \right )^{-1/2}
\left ( 1 + \frac{b_3}{D_3^3}\right ) = 1.31 \;  ,
\ee
which is in perfect agreement with numerical calculations for $c_1$.

The fifth-order corrected root approximants of subsections 10.4 and 
10.5 provide very accurate approximations for the coefficient $c_1$.

\section{Conclusion}

In this paper, we have addressed the problem of extrapolating functions,
which can be found only in the region of asymptotically weak coupling 
parameters, to the limit of infinitely strong coupling parameters. The
main aim has been, assuming that the strong-coupling exponent is known, 
to find the strong-coupling amplitude. Several examples are considered,
for which Pad\'e approximants do not provide good accuracy or are 
not applicable at all. For these problems, we employ the self-similar 
approximation theory, comparing different variants of self-similar 
approximants. Two new classes of self-similar approximants are suggested,
the corrected approximants and iterated approximants.

Comparing the accuracy of different self-similar approximants for 
different problems, we find the following. For the Debye-H\"uckel
function (Sec. 3), the highest accuracy is provided by the iterated 
root approximants. For the quartic anharmonic oscillator (Sec. 4), the 
best are the self-similar factor approximants. For the polymer chain
(Sec. 5), all considered variants of the self-similar approximants 
provide almost the same accuracy. For the oscillating fluid string 
(Sec. 6), the best are the corrected root approximants. For the 
fluctuating fluid membrane (Sec. 7), the corrected root approximants 
are also the most accurate. For strongly interacting fermions (Sec. 8),
the best are the corrected factor approximants. For the one-dimensional 
Bose system (Sec. 9), the iterated root approximants can give the best 
accuracy, though with a large dispersion between the different-order
approximants. For the Bose-Einstein condensation temperature (Sec. 10),
the corrected root approximants give the best accuracy, yielding the 
results that are in very good agreement with the known numerical 
calculations.   

The considered examples illustrate that the self-similar approximants 
provide a versatile tool for extrapolation problems.

\vskip 5mm
{\bf Acknowledgments}

The authors are grateful to E.P. Yukalova for useful discussions. One
of the authors (V.I.Y.) appreciates financial support from the Russian 
Foundation for Basic Research.


\newpage

\begin{thebibliography}{99}

\bibitem{1}
H. Kleinert, {\it Path Integrals} (World Scientific, Singapore, 2006).

\bibitem{2}
V.I. Yukalov, Phys. Rev. A {\bf 42}, 3324 (1990).

\bibitem{3}
V.I. Yukalov, Physica A {\bf 167}, 833 (1990).

\bibitem{4}
V.I. Yukalov, J. Math. Phys. {\bf 32}, 1235 (1991).

\bibitem{5}
V.I. Yukalov, J. Math. Phys. {\bf 33}, 3994 (1992).

\bibitem{6}
V.I. Yukalov, E.P. Yukalova, Physica A {\bf 198}, 573 (1993).

\bibitem{7}
V.I. Yukalov, E.P. Yukalova, Physica A {\bf 206}, 553 (1994).

\bibitem{8}
V.I. Yukalov, E.P. Yukalova, Physica A {\bf 225}, 336 (1996).

\bibitem{9}
V.I. Yukalov, S. Gluzman, Phys. Rev. Lett. {\bf 79}, 333 (1997).

\bibitem{10}
S. Gluzman, V.I. Yukalov, Phys. Rev. E {\bf 58}, 4197 (1998).

\bibitem{11}
V.I. Yukalov, E.P. Yukalova, S. Gluzman, Phys. Rev. A {\bf 58}, 
96 (1998).

\bibitem{12}
V.I. Yukalov, S. Gluzman, Physica A {\bf 273}, 401 (1999).

\bibitem{13}
G.A. Baker, P. Graves-Moris, {\it Pad\'e Approximants} 
(Cambridge University, Cambridge, 1996). 

\bibitem{14}
S. Gluzman, V.I. Yukalov, J. Math. Chem. {\bf 39}, 47 (2006).

\bibitem{15}
V.I. Yukalov, E.P. Yukalova, S. Gluzman, J. Math. Chem. in press. 

\bibitem{16}
V.I. Yukalov, S. Gluzman, Phys. Rev. E {\bf 55}, 6552 (1997).

\bibitem{17}
V.I.Yukalov, S. Gluzman, Phys. Rev. E {\bf 58}, 1359 (1998).

\bibitem{18}
S. Gluzman, V.I. Yukalov, D. Sornette, Phys. Rev. E {\bf 67}, 
026109 (2003)

\bibitem{19}
V.I. Yukalov, S. Gluzman, D. Sornette, Physica A {\bf 328}, 
409 (2003). 

\bibitem{20}
V.I. Yukalov, E.P. Yukalova, Phys. Lett. A {\bf 368}, 341 (2007).

\bibitem{21}
E.P. Yukalova, V.I. Yukalov, S. Gluzman, Ann. Phys. (NY) {\bf 323}, 
3074 (2008).

\bibitem{22}
V.I. Yukalov, E.P. Yukalova, Chaos Solit. Fract. {\bf 14}, 839 (2002).

\bibitem{23}
E.A. Guggenheim, {\it Thermodynamics} (North-Holland, Amsterdam, 1957).

\bibitem{24}
L.D. Landau, E.M. Lifshitz {\it Statistical Physics} 
(Butterworth-Heinemann, Oxford, 2000).

\bibitem{25}
F.T. Hioe, D. McMillen, E.W. Montroll, Phys. Rep. {\bf 43}, 307 (1978).

\bibitem{26}
V.I. Yukalov, E.P. Yukalova, Laser Phys. {\bf 5}, 154 (1995).

\bibitem{27}
M. Muthukumar, B.G. Nickel, J. Chem. Phys. {\bf 80}, 5839 (1984).

\bibitem{28}
M. Muthukumar, B.G. Nickel, J. Chem. Phys. {\bf 86}, 460 (1987).

\bibitem{29}
B. Li, N. Madras, A.D. Sokal, J. Stat. Phys. {\bf 80}, 661 (1995).

\bibitem{30}
U. Seifert, Adv. Phys. {\bf 46}, 13 (1997).

\bibitem{31}
S.F. Edwards, Proc. Roy. Soc. London {\bf 85}, 613 (1965).

\bibitem{32}
H. Kleinert, Phys. Lett. A {\bf 257}, 269 (1999).

\bibitem{33}
B. Kastening, Phys. Rev. E {\bf 66}, 061102 (2002).

\bibitem{34}
B. Kastening, Phys. Rev. E {\bf 73}, 011101 (2006).

\bibitem{35}
G. Gompper, D.M. Kroll, Eur. Phys. Lett. {\bf 9}, 59 (1989).

\bibitem{36}
G.A. Baker, Phys. Rev. C {\bf 60}, 054311 (1999).

\bibitem{37}
W. Ketterle, M.W. Zwierlein, Riv. Nuovo Cimento {\bf 31}, 247 (2008).

\bibitem{38}
J. Carlson, S.Y. Chang, V.R. Pandharipande, K.E. Schmidt, Phys. Rev. Lett.
{\bf 91}, 050401 (2003).

\bibitem{39}
G.E. Astrakharchik, J. Boronat, J. Casulleras, S. Giorgini, Phys. Rev. Lett.
{\bf 93}, 200404 (2004).

\bibitem{40}
E.H. Lieb, W. Liniger, Phys. Rev. {\bf 130}, 1605 (1963).

\bibitem{41}
V.I. Yukalov, M.D. Girardeau, Laser Phys. Lett. {\bf 2}, 375 (2005).

\bibitem{42}
J.O. Andersen, Rev. Mod. Phys. {\bf 76}, 599 (2004)

\bibitem{43}
V.I. Yukalov, Laser Phys. Lett. {\bf 1}, 435 (2004).

\bibitem{44}
B. Kastening, Laser Phys. {\bf 14}, 586 (2004).

\bibitem{45}
B. Kastening, Phys. Rev. A {\bf 69}, 043613 (2004).

\bibitem{46}
B. Kastening, Phys. Rev. A {\bf 70}, 043621 (2004). 

\bibitem{47}
P. Arnold, G. Moore, Phys. Rev. Lett. {\bf 87}, 120401 (2001).

\bibitem{48}
P. Arnold, G. Moore, Phys. Rev. E {\bf 64}, 066113 (2001).

\bibitem{49}
V.A. Kashurnikov, N. Prokofiev, B. Svistunov, Phys. Rev. Lett. {\bf 87}, 
120402 (2001).

\bibitem{50}
N. Prokofiev, B. Svistunov, Phys. Rev. Lett. {\bf 87}, 160601 (2001).

\bibitem{51}
K. Nho, D.P. Landau, Phys. Rev. A {\bf 70}, 053614 (2004).

\bibitem{52}
V.I. Yukalov, Moscow Univ. Phys. Bull. {\bf 31}, 10 (1976).

\bibitem{53}
J.L. Kneur, M.B. Pinto, R.O. Ramos, Phys. Rev. Lett. {\bf 89}, 210403 
(2002).

\bibitem{54}
J.L. Kneur, A. Neveu, M.B. Pinto, Phys. Rev. A {\bf 69}, 053624 (2004).

\bibitem{55}
J.L. Kneur, M.B. Pinto, Phys. Rev. A {\bf 71}, 033613 (2005).
\end{thebibliography}
\end{document}